# Materials Contrast in Piezoresponse Force Microscopy

Sergei V. Kalinin[*]

Materials Science and Technology Division, Oak Ridge National Laboratory,
Oak Ridge, TN 37831

Eugene A. Eliseev

Institute for Problems of Materials Science, National Academy of Science of Ukraine,
3, Krjijanovskogo, 03142 Kiev, Ukraine

Anna N. Morozovska[†]

V. Lashkaryov Institute of Semiconductor Physics, National Academy of Science of Ukraine,
41, pr. Nauki, 03028 Kiev, Ukraine

**Abstract**

Piezoresponse Force Microscopy (PFM) contrast in transversally isotropic material corresponding to the case of $c^+$ - $c^-$ domains in tetragonal ferroelectrics is analyzed using Green's function theory by Felten et al., [J. Appl. Phys. 96, 563 (2004)]. A simplified expression for the PFM signal as a linear combination of relevant piezoelectric constants is obtained. This analysis is extended to piezoelectric material of arbitrary symmetry with weak elastic and dielectric anisotropies. These results provides a framework for interpretation of PFM signals for systems with unknown or poorly known local elastic and dielectric properties, including nanocrystalline materials, ferroelectric polymers, and biopolymers.

[*] Corresponding author, sergei2@ornl.gov
[†] Corresponding author, morozo@i.com.ua

In the last decade, Piezorensponse Force Microscopy (PFM) has emerged as a primary tool for the characterization of ferroelectric and piezoelectric materials on the nanoscale.[1,2,3,4,5] The ability to image ferroelectric domain structures with a nanometer resolution, relative insensitivity to the topographic cross-talk, and the capability to probe local switching behavior necessitate quantitative interpretation of the PFM signal in terms of the relevant material properties. Rigorous calculation of the electromechanical response induced by the biased tip requires the solution of the coupled electromechanical indentation problem, currently available only for the transversally isotropic case.[6,7] This solution is limited to the strong indentation limit, in which the fields generated outside the contact area are neglected. An alternative approach is based on a decoupling approximation, in which the electric field is calculated using the rigid electrostatic model (no piezoelectric coupling), the strain distribution is determined using the constitutive relations for piezoelectric material, and the displacement field is evaluated using the appropriate Green's function for non-piezoelectric solid. The 1D version of this model was originally suggested by Ganpule et al.,[8] to account for the effect of 90° domain walls on PFM image. The electrostatic field was calculated using a 3D model. A 1D approach was later adapted by Agronin et al.,[9] to yield closed-form solutions for the PFM signal. The decoupling approach was extended to 3D by Felten et al.,[10] using an analytical form for the corresponding Green's function. Independently, Scrymgeour and Gopalan[11] have used finite element method to calculate the PFM signals across the domain walls. The advantage of the decoupled 3D models is that the PFM signal can be determined for an arbitrary electric fields, and the PFM response can be calculated for various microstructural elements. However, existing solutions are extremely cumbersome and require numerical calculations.

Here, we analyze PFM contrast in the transversally isotropic material using the Green's function theory suggested by Felten et al.[10] The closed-form expression for the PFM signal, including relative contributions of the individual elements of the piezoelectric constant tensor, elastic properties, and dielectric anisotropy, are derived. This analysis is extended to the anisotropic piezoelectric with weak elastic and dielectric anisotropies.

The Green's function approach is based on the (1) calculation of the electric field for rigid dielectric ($d_{ijk} = e_{ijk} = 0$), (2) calculation of stress field $X_{ij} = e_{kij}E_k$ in piezoelectric material, and (3) calculation of the mechanical displacement using Green's function for non-piezoelectric elastic half-plane. This approximation significantly simplifies the problem and in particular allows the effective symmetries of the elastic, dielectric, and piezoelectric properties to be varied independently. Shown in Fig. 1 are crystallographic orientation dependence for the effective dielectric, piezoelectric, and elastic properties for $BaTiO_3$ and $LiNbO_3$.[12] In particular, we note that the dielectric and particularly the elastic properties described by the positively defined second- and fourth-rank tensors are necessarily more isotropic then piezoelectric properties described by a third-rank tensor. Hence, in many cases the elastic and dielectric properties can be approximated as isotropic.

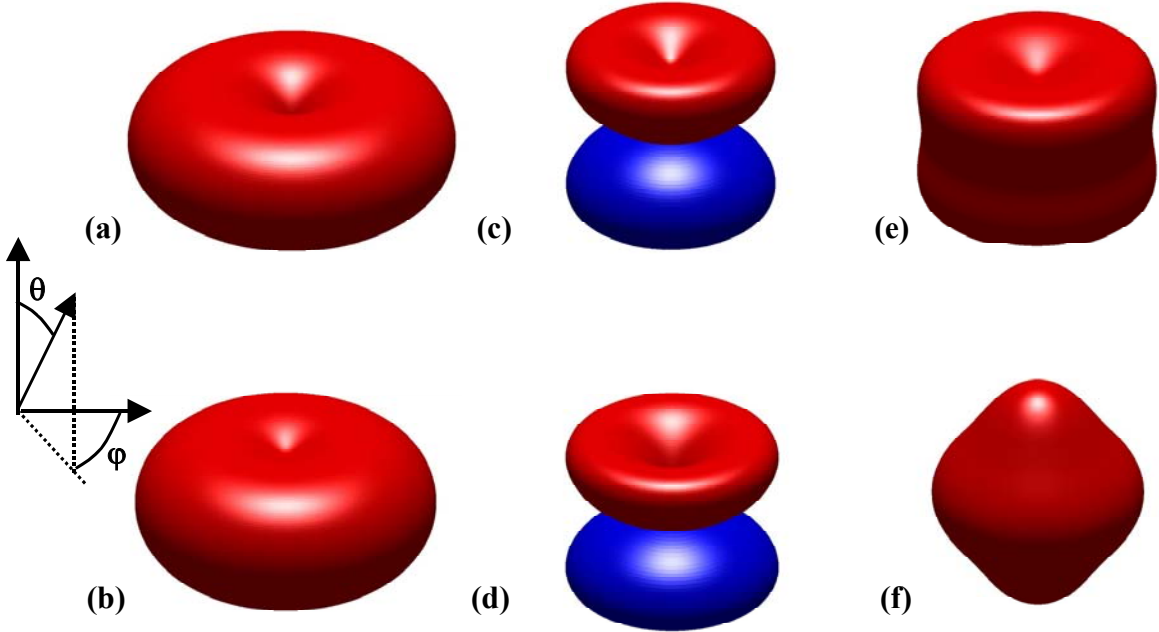

**Fig. 1.** Crystallographic orientation dependence of effective dielectric constant in z-direction (a,b), effective longitudinal piezoelectric constant (c,d) and effective Young's modulus (e,f) for $BaTiO_3$ (a,c,e) and $LiNbO_3$ (b,d,f).

For transversally isotropic material, the tip-induced electric field can be determined using image charge models.[13,14] The potential produced by the point charge, $Q$, at a distance, $d$, above the surface, is

$$V_i(\rho, z) = \frac{Q}{2\pi\varepsilon_0(\kappa+1)} \frac{1}{\sqrt{\rho^2 + (z/\gamma + d)^2}}, \quad (1)$$

where $\rho$ and $z$ are the radial and vertical coordinate, $\kappa = \sqrt{\varepsilon_{33}\varepsilon_{11}}$ is the effective dielectric constant, and $\gamma = \sqrt{\varepsilon_{33}/\varepsilon_{11}}$ is the dielectric anisotropy factor [Fig. 2(a)].

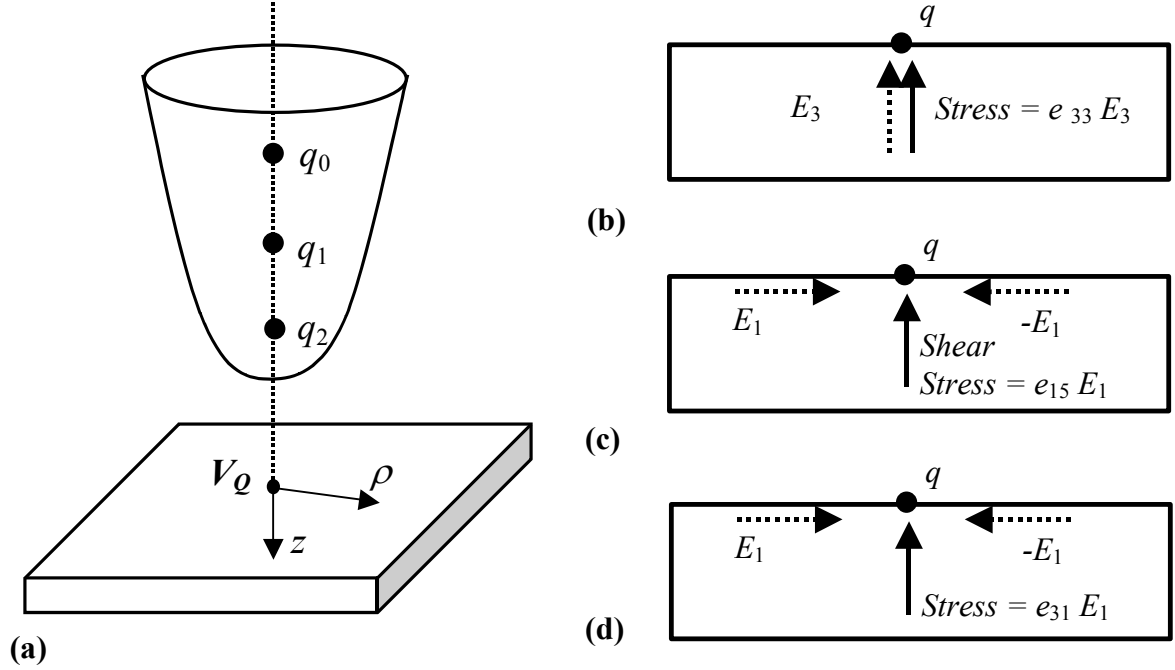

**Fig. 2.** (a) Tip representation using image charge distribution in the PFM experiment. Schematics of contributions of (b) $e_{33}$, (c) $e_{15}$ and (d) $e_{31}$ to the PFM signal.

The displacement field can be calculated using the Green's function approach suggested by Felten et al.[10] The displacement vector $u_i(\mathbf{x})$ at position $\mathbf{x}$ is

$$u_i(\mathbf{x}) = \int_{x_3=0}^{\infty}\int_{x_2=-\infty}^{\infty}\int_{x_1=-\infty}^{\infty} \frac{\partial G_{ij}(\mathbf{x},\boldsymbol{\xi})}{\partial \xi_l} E_k(\boldsymbol{\xi})d\boldsymbol{\xi}\, e_{kjl} \qquad (2)$$

where $\boldsymbol{\xi}$ is the coordinate system related to the material, $e_{kjl}$ are the strain piezoelectric coefficients ($e_{kij} = d_{klm}c_{lmij}$) and the Einstein summation convention is used. $E_k(\boldsymbol{\xi})$ is the electric field produced by the probe. For most ferroelectric perovskites, the symmetry of the elastic properties can be approximated as cubic (anisotropy of the elastic properties is much smaller than that of the dielectric and piezoelectric properties) and therefore isotropic approximation is used. The Green's function for isotropic semi-infinite half-plane is given by Landau and Lifshitz[15] and depends on the Young's modulus, $Y$, and the Poisson ratio, $\nu$. After lengthy manipulations (see Appendices A-C), the Eq. (2) is integrated analytically to yield the transverse displacement of the surface ($z = 0$) at the position of the tip, i.e., the vertical PFM signal, as

$$u_3(\rho) = \frac{Q}{2\pi\varepsilon_0} \frac{1+\nu}{Y} \frac{1}{\sqrt{\rho^2 + d^2}} (e_{31} f_1(\gamma) + e_{15} f_2(\gamma) + e_{33} f_3(\gamma)), \qquad (3)$$

and $u_1(0) = u_2(0) = 0$. Hereinafter we use $e_{31} \equiv e_{311}$, $e_{33} \equiv e_{333}$, $e_{15} \equiv e_{113}$ in Voigt representation when possible. The functions $f_i(\gamma)$ that determine the contributions of the piezoelectric constants $e_{in}$ to the overall signal depend only on the dielectric anisotropy, $\gamma$, as

$$f_1(\gamma) = \left( \frac{\gamma}{(1+\gamma)^2} - \frac{(1-2\nu)}{(1+\gamma)} \right) \qquad (4)$$

$$f_2(\gamma) = -\frac{2\gamma^2}{(1+\gamma)^2} \qquad (5)$$

$$f_3(\gamma) = -\left( \frac{\gamma}{(1+\gamma)^2} + \frac{(1-2\nu)}{1+\gamma} \right) \qquad (6)$$

The contributions of different piezoelectric constants to the overall displacement are shown schematically in Fig. 2. The normal component of the electric field is related to the vertical stress component by piezoelectric constant $e_{33}$ [Fig. 2 (b)]. The second contribution to the response originates from the lateral component of the electric fields related to the shear stress component by piezoelectric constant $e_{15}$ [Fig. 2 (c)]. Finally, the constant $e_{31}$ relates the stress in z-direction to the normal field component [Fig. 2 (d)].

Shown in Fig. 3 (a,b) are the functions $f_i(\gamma)$ that determine contributions of the piezoelectric constants $e_{nm}$ to the overall signal. For the majority of ferroelectric oxides $\gamma \approx 0.3-1$, while $\gamma \approx 10.5$ and 2.3 for Rochelle salt and triglycine sulfate respectively. The function $f_3(\gamma)$ rapidly decays with $\gamma$, indicating the decreasing contribution of $e_{33}$ to the signal. The signal decreases for low compressibility materials (high ν). Conversely, $f_2(\gamma)$ that determines contribution of constant $e_{15}$ increases with $\gamma$. Finally, $f_1(\gamma)$ has a maximum and is much smaller then $f_3(\gamma)$ and $f_2(\gamma)$.

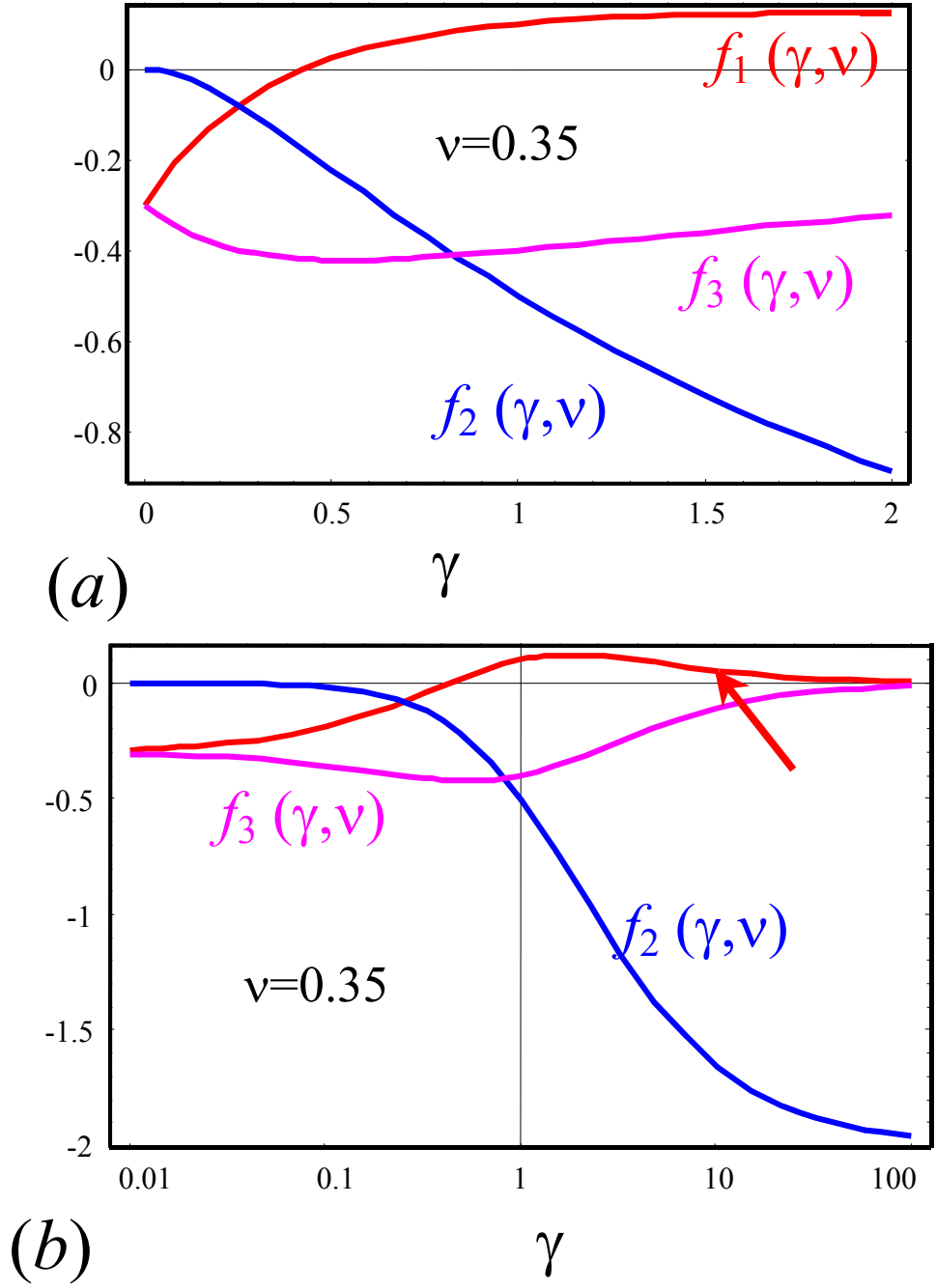

**Fig. 3.** (a) Plot of functions $f_i(\gamma,\nu)$ for $\nu = 0.35$ in the $(0,1)$ interval corresponding to most ferroelectric oxides. (b) Plot in the logarithmic scale in the $(10^{-2}, 10^2)$ interval illustrating the asymptotic behavior of the dissimilar contributions with dielectric anisotropy.

To calculate the PFM signal, we note the similarity between Eq. (4) and Eq. (1) for $z = 0$. Hence, we obtain

$$u_3(0) = V_Q \frac{1+\nu}{Y}(e_{31}f_1(\gamma) + e_{15}f_2(\gamma) + e_{33}f_3(\gamma)), \qquad (7)$$

where $V_Q = V_i(\rho, z = 0)$. Thus, the PFM response is proportional to the potential on the surface induced by the charge. This is also true for charges located on the same line along the surface normal, i.e., for the sphere and line charge approximations for the tip. Hence, the response depends only on the maximal surface potential induced by the tip, and does not depend on the exact charge distribution, in agreement with previous results.[13]

The electromechanical responses calculated using Eq. (7) and the exact theory[6] are shown in Table I. In the first column, compared are the elastic modulus $C_1^*/2\pi$ and Young's modulus in the z-direction $1/s_{33}$.[12]

Table I

Materials Properties and Calculated Responses

| Material | Elastic constants [GPa] | | Dielectric anisotropy | Piezoelectric properties [pm/V] | | Approximate solution [pm/V] | | |
|---|---|---|---|---|---|---|---|---|
| | $C_1^*/2\pi$ | $1/s_{33}$ | $\gamma$ | $d_{33}$ | $C_3^*/C_1^*$ | $d_{33}^*$ | $d_{33}^*(\gamma=1)$ | $d_{33}^*(d)$ |
| BaTiO$_3$ | **128.3** | 63.67 | 0.24 | 85.59 | **38.2** | 29.7 | 130.3 | 55.2 |
| LiNbO$_3$ ‡ | **205.95** | 202.55 | 0.6 | 6.31 | **11.63** | 10.8 | 15.2 | 14.0 |
| LiTaO$_3$ | **248.3** | 228.1 | 0.94 | 8.328 | **11.28** | 11.1 | 11.4 | 11.2 |
| PZT6B | **114.5** | 107.2 | 0.98 | 74.94 | **71.11** | 64.0 | 64.3 | 72.5 |

Note the good agreement between these values for all materials excluding BaTiO$_3$ in which the anisotropy of mechanical properties is significant. In the second column, listed are coefficients of dielectric anisotropy. In the third column, the values of piezoelectric constant are compared, $d_{33}$, and the electromechanical response from exact theory, $C_3^*/C_1^*$. Finally, in the last column we tabulate the values of the electromechanical response $d_{33}^* = u_3(0)/V_Q$ calculated from Eq. (7) for $Y = C_1^*/2\pi$ and $\nu = 0.35$, the electromechanical response for zero dielectric anisotropy $d_{33}^*(\gamma = 1)$, and the electromechanical response, $d_{33}^*(d)$, calculated for the values of $e_{im} = d_{in}c_{nm}$, where the material is assumed to be elastically isotropic. In this latter case when strain piezoelectric constants $d_{in}$ are used, the response is independent of the Young's modulus and is determined only by the Poisson ratio.[10] Note that for materials with weak ($\gamma \approx 1$) and moderate ($\gamma \approx 0.6$) dielectric anisotropies the approximate PFM response from Eq. (7) is in a very good agreement with exact result, validating the use of this model. The relative analysis of different contributions in Eq. (7) suggests that the response is dominated by $e_{33}$ and $e_{15}$ terms, while $e_{31}$ provides only a minor contribution [e.g., for BTO the relative contributions of $(e_{31}, e_{33}, e_{15})$ are $(0.06, 44, 50)$, for LiNbO$_3$ $(0.01, 35, 64)$, and for PZT 6B $(0.03, 54, 43)$].

---

‡ Although components $e_{22} = -e_{21} = -e_{16}$ of piezoelectric tensor are nonzero for LiNbO$_3$ and LiTaO$_3$, their contribution to the displacement below the tip [Eqs. (7) and (8)] are zero due to the transversal isotropy of the system.

From the data in Table I, certain material can be approximated as both elastically and dielectrically isotropic. Here we extend Eq. (7) to the fully anisotropic piezoelectric solids with a weak dielectric anisotropy ($\varepsilon_{ij} \approx \delta_{ij}\kappa\varepsilon_0$). In this case, after lengthy integrations (Appendix C) the components of the surface displacement related to vertical and lateral PFM signals [16,17] are found from Eqs. (1,2) as $u_i(\mathbf{x}) = V_Q(0)U_{ijlk}(\mathbf{x})e_{kjl}(1+\nu)/Y$, where tensor $U_{ijlk}(0)$ is symmetrical on the transposition of the indexes $j$ and $l$. In Voigt notation, the displacements are

$$u_1(0) = V_Q(0)\frac{1+\nu}{Y}(U_{111}e_{11} + U_{121}e_{12} + U_{131}e_{13} + U_{153}e_{35} + U_{162}e_{26}) \tag{8a}$$

$$u_2(0) = V_Q(0)\frac{1+\nu}{Y}(U_{121}e_{21} + U_{111}e_{22} + U_{131}e_{23} + U_{153}e_{34} + U_{162}e_{16}) \tag{8b}$$

$$u_3(0) = V_Q(0)\frac{1+\nu}{Y}(U_{131}(e_{31} + e_{32}) + U_{333}e_{33} + U_{351}(e_{24} + e_{15})) \tag{8c}$$

The non-zero elements of tensor $U_{i\alpha k}$ are:

$$U_{111} = -\left(\frac{7 + 6(1-2\nu)}{32}\right), \qquad U_{121} = \left(\frac{3 - 2(1-2\nu)}{32}\right), \tag{9a-b}$$

$$U_{131} = \left(\frac{1-2(1-2\nu)}{8}\right), \qquad U_{162} = -\left(\frac{5+2(1-2\nu)}{16}\right), \qquad (9\text{c-d})$$

$$U_{153} = -\frac{3}{4}, \quad U_{351} = -\frac{1}{4}, \quad U_{333} = -\left(\frac{1+2(1-2\nu)}{4}\right). \qquad (9\text{e-g})$$

Thus, Eq. (8) provides an approximate description of the PFM response for the material of arbitrary symmetry and crystallographic orientation. The detailed analysis of PFM contrast in the anisotropic case is reported elsewhere.[18]

To summarize, a simplified expression for the PFM signal as a linear combination of the relevant piezoelectric constant is obtained for transversally isotropic dielectric medium with weak elastic anisotropy. The solution can be readily extended to an arbitrary tip model the response is shown to be proportional to the potential induced by the tip on the surface and is independent of the exact image charge distribution. This analysis is extended to piezoelectric materials of an arbitrary symmetry with weak elastic and dielectric anisotropies. The calculated response is dependent on either (a) the piezoelectric strain constants, $e_{ijk}$, of the materials and elastic properties $(Y,\nu)$ that can be determined from the indentation experiments or (b) piezoelectric stress constants, $d_{ijk}$, and Poisson modulus, $\nu$, only. This analysis provides a framework for the interpretation of the PFM signal in systems with unknown or poorly known local elastic properties, including nanoferroelectrics, ferroelectric polymers, and biopolymers.

### APPENDIX A. The statement of the problem.

For linear piezoelectric material, the relationship between strain $U_{ij}$, displacement $D_i$, stress $X_{kl}$ and electric field $E_m$ is

$$U_{ij} = s_{ijkl} X_{kl} + E_m d_{mij}, \qquad (A.1)$$

$$D_i = d_{ijk} X_{jk} + \varepsilon_{im} E_m. \qquad (A.2)$$

The relative contributions of different terms in Eqs. (A.1, 2) are estimated using simple model. For the contact radius of ~5 nm, corresponding to the indentation force of ~100 nN and tip radius of curvature of 50 nm, the average stress below the tip is 1.27 GPa. The electric field for tip bias of 10 V is $2 \cdot 10^9$ V/m. For typical elastic compliance of order of $10^{-11}$ m$^2$/N and piezoelectric constant 50 pm/V, the first term in Eq. (A.1) is 0.013 and the second is 0.1. In Eq. (A.2) for direct effect for dielectric constant of 100, the first term is 0.064 and second is 1.77. From this simple estimate, the dielectric term dominates Eq. (A.2), thus justifying the use of rigid dielectric approximation for calculating the electric field in the material. This analysis is corroborated by the exact solution for transversally isotropic case.

Let us multiply Eq. (A.1) by the matrix of elastic stiffness $c_{ijpq}$ ($\hat{c} \cdot \hat{s} = \hat{I}$):

$$c_{ijpq} U_{ij} = c_{ijpq} s_{ijkl} X_{kl} + E_m d_{mij} c_{ijpq}, \tag{A.3}$$

$$c_{ijpq} U_{ij} - E_m d_{mij} c_{ijpq} = X_{pq}, \tag{A.4}$$

$$X_{pq} = c_{pqij} U_{ij} - E_m e_{mpq}, \qquad e_{mpq} = d_{mij} c_{ijpq}. \tag{A.5}$$

The tensor $X_{pq}$ must satisfy the equilibrium conditions $\partial X_{pq}/\partial x_p = 0$, thus $c_{pqij} \partial U_{ij}/\partial x_p = \partial(E_m e_{mpq})/\partial x_p$. Therefore the force acting in the bulk of the system has the density $F_k = -\partial(E_i e_{ijk})/\partial x_j$. The pressure acting on the sample surface $z = 0$ has the view $P_k = -E_i e_{i3k}$. The displacement is:

$$u_i(\mathbf{x}) = -\int_0^\infty \int_{-\infty}^\infty \int_{-\infty}^\infty G_{ij}(\mathbf{x}, \xi_1, \xi_2, \xi_3) \frac{\partial E_k(\xi_1, \xi_2, \xi_3)}{\partial \xi_l} e_{klj} d\xi_1 d\xi_2 d\xi_3 -$$
$$- \int_{-\infty}^\infty \int_{-\infty}^\infty G_{ij}(\mathbf{x}, \xi_1, \xi_2, \xi_3 = 0) E_k(\xi_1, \xi_2, \xi_3 = 0) e_{k3j} d\xi_1 d\xi_2 \tag{A.6}$$

Integration in parts in Eq. (A.6) leads to the following expression:

$$u_i(\mathbf{x}) = \int_0^\infty \int_{-\infty}^\infty \int_{-\infty}^\infty \frac{\partial G_{ij}(\mathbf{x}, \xi_1, \xi_2, \xi_3)}{\partial \xi_l} E_k(\xi_1, \xi_2, \xi_3) e_{klj} d\xi_1 d\xi_2 d\xi_3 \tag{A.7}$$

Here Green's tensor component $G_{ij}(\mathbf{x}, \xi)$ determines the displacement $u_i(\mathbf{x})$ at the point $\mathbf{x}$ under the point force $\mathbf{F}$ component "$j$" applied at the point $\xi$. It is defined by the relation $u_i = G_{ij} F_j$. At the surface ($x_3 = 0$) Green's tensor is as follows from Ref. 15:

$$G_{ij}(x_1,x_2,x_3=0,\boldsymbol{\xi})=\begin{cases} \dfrac{1+\nu}{2\pi Y}\left[\dfrac{\delta_{ij}}{R}+\dfrac{(x_i-\xi_i)(x_j-\xi_j)}{R^3}+\dfrac{1-2\nu}{R+\xi_3}\left(\delta_{ij}-\dfrac{(x_i-\xi_i)(x_j-\xi_j)}{R(R+\xi_3)}\right)\right] & i,j\neq 3 \\[6pt] \dfrac{(1+\nu)(x_i-\xi_i)}{2\pi Y}\left(\dfrac{-\xi_3}{R^3}-\dfrac{(1-2\nu)}{R(R+\xi_3)}\right) & i=1,2\ \text{and}\ j=3 \\[6pt] \dfrac{(1+\nu)(x_j-\xi_j)}{2\pi Y}\left(\dfrac{-\xi_3}{R^3}+\dfrac{(1-2\nu)}{R(R+\xi_3)}\right) & j=1,2\ \text{and}\ i=3 \\[6pt] \dfrac{1+\nu}{2\pi Y}\left(\dfrac{2(1-\nu)}{R}+\dfrac{\xi_3^2}{R^3}\right) & i=j=3 \end{cases}$$

(A.8)

Here the following designation is introduced $R=\sqrt{(x_1-\xi_1)^2+(x_2-\xi_2)^2+\xi_3^2}$. It is seen that tensor $G_{ij}$ is nonsymmetrical, since $G_{13}(\boldsymbol{x};\boldsymbol{\xi})\neq G_{31}(\boldsymbol{x};\boldsymbol{\xi})$ and $G_{23}(\boldsymbol{x};\boldsymbol{\xi})\neq G_{32}(\boldsymbol{x};\boldsymbol{\xi})$, while $G_{12}(\boldsymbol{x};\boldsymbol{\xi})=G_{21}(\boldsymbol{x};\boldsymbol{\xi})$.

Since the electric field distribution for the general case of material with arbitrary dielectric anisotropy can be presented in the form of Fourier integrals (see Appendix B) it is natural to turn to this representation. It is seen that Green's tensor components (A.8) depend only on the differences $x_1-\xi_1$ and $x_2-\xi_2$, therefore we can write:

$$G_{ij}(x_1,x_2,x_3=0,\boldsymbol{\xi})=\frac{1}{2\pi}\int_{-\infty}^{\infty}dk_1\int_{-\infty}^{\infty}dk_2\,\exp(-ik_1(x_1-\xi_1)-ik_2(x_2-\xi_2))\cdot\widetilde{G}_{ij}(k_1,k_2,\xi_3)$$

(A.9)

where Fourier image is

$$\widetilde{G}_{ij}(k_1,k_2,\xi)=\frac{1+\nu}{2\pi Y}\frac{\exp(-\xi k)}{k}\left(2\delta_{ij}-\frac{k_ik_j}{k^2}(\xi k+2\nu)\right),\quad i,j\neq 3 \tag{A.10a}$$

$$\widetilde{G}_{i3}(k_1,k_2,\xi)=-\frac{1+\nu}{2\pi Y}\cdot\frac{ik_i\exp(-\xi k)}{k^2}(\xi k+(1-2\nu)),\quad i=1,2 \tag{A.10b}$$

$$\widetilde{G}_{3j}(k_1,k_2,\xi)=-\frac{1+\nu}{2\pi Y}\cdot\frac{ik_j\exp(-\xi k)}{k^2}(\xi k-(1-2\nu)),\quad j=1,2 \tag{A.10c}$$

$$\widetilde{G}_{33}(k_1,k_2,\xi)=\frac{1+\nu}{2\pi Y}\frac{\exp(-\xi k)}{k}(2(1-\nu)+\xi k) \tag{A.10d}$$

Hereinafter we use the designation $k\equiv\sqrt{k_1^2+k_2^2}$. The Fourier representation $\widetilde{G}_{ij,l}(k_1,k_2,\xi)$ of the Green's function gradient $\partial G_{ij}/\partial\xi_l$ can be found from Eq. (A.9), namely:

$$\widetilde{G}_{ij,l}(k_1,k_2,\xi) \equiv \begin{cases} ik_l \widetilde{G}_{ij}(k_1,k_2,\xi), & l=1,2 \\ \dfrac{\partial}{\partial \xi}\widetilde{G}_{ij}(k_1,k_2,\xi), & l=3 \end{cases} \qquad (A.11)$$

Then using the Fourier image of the electric field distribution (see Appendix B for details):

$$E_k(\xi_1,\xi_2,\xi_3) = \frac{1}{2\pi}\int_{-\infty}^{\infty} d\widetilde{k}_1 \int_{-\infty}^{\infty} d\widetilde{k}_2 \exp(-i\widetilde{k}_1\xi_1 - i\widetilde{k}_2\xi_2)\cdot \widetilde{E}_k(\widetilde{k}_1,\widetilde{k}_2,\xi_3) \qquad (A.12)$$

we can rewrite Eq. (A.7) as follows:

$$u_i(x_1,x_2) = \int_{-\infty}^{\infty} dk_1 \int_{-\infty}^{\infty} dk_2 \exp(-ik_1 x_1 - ik_2 x_2)\cdot \int_{0}^{\infty} d\xi\, \widetilde{G}_{ij,l}(k_1,k_2,\xi)\widetilde{E}_k(k_1,k_2,\xi)e_{klj} \qquad (A.13)$$

It is seen that the displacement also represents a threefold integral in this representation, which has the much simpler structure than initial one Eq. (A.7) in many cases. In order to obtain the closed form solution we need the explicit form of the electric field image.

### APPENDIX B. Electric field.

Next we find the representation for the electric field induced by a point charge $Q$ located at the distance $d$ above the surface of the anisotropic half-space with dielectric permittivity tensor $\varepsilon_{ij}$. This field potential $V(\mathbf{x})$ (at $x_3 \geq 0$) and $V_0(\mathbf{x})$ (at $x_3 < 0$) can be obtained from the solution of the Laplace's equations:

$$\begin{cases} \varepsilon_0 \varepsilon_{ij}\dfrac{\partial^2}{\partial x_i \partial x_j} V(\mathbf{x}) = 0, & x_3 \geq 0 \\ \varepsilon_0 \Delta V_0(\mathbf{x}) = -Q\cdot\delta(x_3+d)\delta(x_2)\delta(x_1), & x_3 < 0 \end{cases} \qquad (B.1)$$

$$\varepsilon_{3j}\left.\frac{\partial V(\mathbf{x})}{\partial x_j}\right|_{x_3=0} = \left.\frac{\partial V_0(\mathbf{x})}{\partial x_3}\right|_{x_3=0}, \quad V(x_3=0) = V_0(x_3=0)$$

Let us introduce Fourier images

$$V(\mathbf{x}) = \frac{1}{2\pi}\int_{-\infty}^{\infty} dk_1 \int_{-\infty}^{\infty} dk_2 \exp(-ik_1 x_1 - ik_2 x_2)\cdot \widetilde{V}(k_1,k_2,x_3),$$

$$V_0(\mathbf{x}) = \frac{1}{2\pi}\int_{-\infty}^{\infty} dk_1 \int_{-\infty}^{\infty} dk_2 \exp(-ik_1 x_1 - ik_2 x_2)\cdot \widetilde{V}_0(k_1,k_2,x_3)$$

The equations for Fourier images can be obtained from (B.1) as follows:

$$\begin{cases} \left(\varepsilon_{33}\dfrac{\partial^2}{\partial x_3^2} - 2i(\varepsilon_{31}k_1 + \varepsilon_{32}k_2)\dfrac{\partial}{\partial x_3} - \varepsilon_{11}k_1^2 - 2\varepsilon_{12}k_1k_2 - \varepsilon_{22}k_2^2\right)\tilde{V} = 0, \quad x_3 \geq 0 \\ \left(\dfrac{\partial^2}{\partial x_3^2} - k_1^2 - k_2^2\right)\tilde{V}_0 = -\dfrac{Q}{2\pi\varepsilon_0}\cdot\delta(x_3 + d), \quad x_3 < 0 \\ \left(-i\varepsilon_{31}k_1 - i\varepsilon_{32}k_2 + \varepsilon_{33}\dfrac{\partial}{\partial x_3}\right)\tilde{V} = \dfrac{\partial}{\partial x_3}\tilde{V}_0(x_3 = 0), \quad \tilde{V}(x_3 = 0) = \tilde{V}_0(x_3 = 0) \end{cases} \quad (B.2)$$

We used Dirac-delta representation $\delta(x) = \dfrac{1}{2\pi}\int\limits_{-\infty}^{\infty} dx \exp(-ikx)$.

Let us find the solution of (B.2) at $x_3 < 0$ in the form

$\tilde{V}_0(x_3) = C_0(k_1, k_2)\exp(-k|x_3 + d|) + A_0(k_1, k_2)\exp(kx_3)$, where $k \equiv \sqrt{k_1^2 + k_2^2}$, the constant $C_0(k_1, k_2)$ satisfies the equation:

$(\partial^2/\partial x_3^2 - k^2)\tilde{V}_0(x_3) = -2kC_0(k_1, k_2)\exp(-k\cdot|x_3 + d|)\delta(x_3 + d) = -\delta(x_3 + d)Q/2\pi\varepsilon_0$.

Thus $C_0 = Q/(4\pi\varepsilon_0 k)$.

Let us find the solution of (B.2) at $x_3 \geq 0$ in the form $\tilde{V}(x_3) = B(k_1, k_2)\exp(-\lambda x_3)$, where the characteristic equation has the form

$\varepsilon_{33}\lambda^2 + 2i(\varepsilon_{31}k_1 + \varepsilon_{32}k_2)\lambda - \varepsilon_{11}k_1^2 - 2\varepsilon_{12}k_1k_2 - \varepsilon_{22}k_2^2 = 0$. It is easy to obtain that its root with positive real part has the view:

$$\lambda(\mathbf{k}) = \dfrac{-i(\varepsilon_{31}k_1 + \varepsilon_{32}k_2) + \sqrt{(\varepsilon_{11}\varepsilon_{33} - \varepsilon_{31}^2)k_1^2 - 2(\varepsilon_{31}\varepsilon_{32} - \varepsilon_{12}\varepsilon_{33})k_1k_2 + (\varepsilon_{22}\varepsilon_{33} - \varepsilon_{32}^2)k_2^2}}{\varepsilon_{33}} \quad (B.3)$$

The square root in Eq. (B.3) is real for any real $\mathbf{k} = (k_1, k_2)$ since the matrix of static susceptibility $\varepsilon_{ij}$ is positively defined.

For the constants $B(k_1, k_2)$ and $A_0(k_1, k_2)$ we obtain the system of equations:

$$\begin{cases} B(k_1, k_2) = \dfrac{Q}{4\pi\varepsilon_0 k}\exp(-kd) + A_0(k_1, k_2) \\ -(i\varepsilon_{31}k_1 + i\varepsilon_{32}k_2 + \varepsilon_{33}\lambda(\mathbf{k}))B(k_1, k_2) = -\dfrac{Q}{4\pi\varepsilon_0}\exp(-kd) + kA_0(k_1, k_2) \end{cases} \quad (B.4)$$

It is easy to obtain from (B.4) that

$$A_0(k_1,k_2) = \frac{Q\exp(-kd)}{4\pi\varepsilon_0 k}\left(\frac{k-(i\varepsilon_{31}k_1+i\varepsilon_{32}k_2+\varepsilon_{33}\lambda(\mathbf{k}))}{k+(i\varepsilon_{31}k_1+i\varepsilon_{32}k_2+\varepsilon_{33}\lambda(\mathbf{k}))}\right),$$
$$B(k_1,k_2) = \frac{2Q\exp(-kd)}{4\pi\varepsilon_0(k+(i\varepsilon_{31}k_1+i\varepsilon_{32}k_2+\varepsilon_{33}\lambda(\mathbf{k})))}.$$
(B.5)

Thus

$$\tilde{V}(k_1,k_2,x_3) = \frac{2Q\exp(-kd-\lambda(\mathbf{k})x_3)}{4\pi\varepsilon_0(k+(i\varepsilon_{31}k_1+i\varepsilon_{32}k_2+\varepsilon_{33}\lambda(\mathbf{k})))} \quad (B.6)$$

and original

$$V(\mathbf{x}) = \frac{1}{2\pi}\int_{-\infty}^{\infty}dk_1\int_{-\infty}^{\infty}dk_2 \frac{2Q\exp(-kd-\lambda(\mathbf{k})x_3)\exp(-ik_1x_1-ik_2x_2)}{4\pi\varepsilon_0\left(k+\sqrt{(\varepsilon_{11}\varepsilon_{33}-\varepsilon_{31}^2)k_1^2-2(\varepsilon_{31}\varepsilon_{32}-\varepsilon_{12}\varepsilon_{33})k_1k_2+(\varepsilon_{22}\varepsilon_{33}-\varepsilon_{32}^2)k_2^2}\right)}$$
(B.7)

Allowing for Eq. (B.7), the Fourier representation $\tilde{E}_k(k_x,k_y,x_3)$ of the electric field $E_k(\mathbf{x})=-\partial V(\mathbf{x})/\partial x_k$ can be easily found as

$$\tilde{E}_i(k_1,k_2,x_3) \equiv \begin{cases} ik_i\tilde{V}(k_1,k_2,x_3), & i=1,2 \\ -\dfrac{\partial}{\partial x_3}\tilde{V}(k_1,k_2,x_3), & i=3 \end{cases} \quad (B.8)$$

The general expression (B.7) allowing for Eq. (B.3) can be essentially simplified in the case of the transversally isotropic material ($\varepsilon_{ij}=\varepsilon_{ii}\delta_{ij}$, $\varepsilon_{11}=\varepsilon_{22}\neq\varepsilon_{33}$).

## APPENDIX C. Displacement calculations.

Since piezoelectric tensor $e_{klj}$ is symmetrical on the indexes $l$ and $j$ transposition it is natural to rewrite Eq. (A.7) as

$$u_i(\mathbf{x}) = W_{ijlk}(\mathbf{x})e_{klj} \quad (C.1)$$

where tensor $W_{ijlk}(\mathbf{x})$ has the view

$$W_{ijlk}(\mathbf{x}) = \int_0^{\infty}d\xi_3\int_{-\infty}^{\infty}d\xi_3\int_{-\infty}^{\infty}d\xi_1\frac{1}{2}\left(\frac{\partial G_{ij}(\mathbf{x},\boldsymbol{\xi})}{\partial \xi_l}+\frac{\partial G_{il}(\mathbf{x},\boldsymbol{\xi})}{\partial \xi_j}\right)E_k(\boldsymbol{\xi}). \quad (C.2)$$

$$W_{ijlk}(\mathbf{x}) = \int_{-\infty}^{\infty}dk_1\int_{-\infty}^{\infty}dk_2\exp(-ik_1x_1-ik_2x_2)\cdot\int_0^{\infty}d\xi\frac{1}{2}\left(\tilde{G}_{ij,l}(k_1,k_2,\xi)+\tilde{G}_{il,j}(k_1,k_2,\xi)\right)\tilde{E}_k(k_1,k_2,\xi)$$

$W_{ijlk}(\mathbf{x})$ is symmetrical only on the indexes $l$ and $j$ transposition ($W_{ijlk}(\mathbf{x})\equiv W_{iljk}(\mathbf{x})$) and thus has 54 nontrivial components in general case.

When integrating in Eq. (C.2) it is convenient to turn to the polar coordinates $k_1 \equiv k\cos(\varphi)$, $k_2 \equiv k\sin(\varphi)$. The integration on $\xi$ and $k$ can be done analytically, since the expressions (A.10), (A.11), (B.6) and (B.8) elementary depend on these coordinates. So, Eq. (C.2) is reduced to the one-fold integral on $\varphi$ which can be expressed in terms of elliptic integrals for general dielectric anisotropy. For the case of the transversely isotropic media these integrals were taken in the elementary functions (see Appendix D).

Hereinafter we consider the results of $W_{ijlk}(\mathbf{x})$ integration for the case of material with weak dielectrically anisotropy ($\varepsilon_{ij} \approx \kappa \delta_{ij}$). The designations $x_1 = x$, $x_2 = y$, $\rho^2 = x^2 + y^2$ and $a = \sqrt{x^2 + y^2 + d^2}$ are introduced for clearness. It should be noted that the results below can be easily generalized for transversely isotropic material.

It is easy to show that different $W_{ijlk}(\mathbf{x})$ that contains indexes "1" or/and "2" can be obtained one from another by simultaneous permutation of indexes "1" $\leftrightarrow$ "2" and coordinates x $\leftrightarrow$ y, e.g.: $W_{1111}(x,y) \equiv W_{2222}(y,x)$, $W_{1223}(x,y) \equiv W_{2113}(y,x)$, $W_{2211}(x,y) \equiv W_{1122}(y,x)$, $W_{2311}(x,y) \equiv W_{1322}(y,x)$, $W_{2333}(x,y) \equiv W_{1333}(y,x)$. Therefore the number of nontrivial components of $W_{ijlk}(\mathbf{x})$ is reduced to 28, which is listed below.

Components determining the displacement component $u_1$ are the following:

$$W_{1111}(x,y) = -\frac{Q}{2\pi\varepsilon_0(\kappa+1)} \frac{1+\nu}{2\pi Y} \frac{\pi}{4a(a+d)^2 \rho^4} \times$$
$$\times \begin{pmatrix} d(5a+2d)x^4 + 2(4a^2 - ad + 4d^2)x^2 y^2 + a(2a+5d)y^4 + \\ + 2(1-2\nu)(d(a+2d)x^4 + 6ad\,x^2 y^2 + a(2a+d)y^4) \end{pmatrix}$$
(C.3a)

$$W_{1112}(x,y) = \frac{Q}{2\pi\varepsilon_0(\kappa+1)} \frac{1+\nu}{2\pi Y} \frac{\pi x y}{2a(a+d)^3 \rho^2} \times$$
$$\times \left((4a+d)x^2 + (a+4d)y^2 + 2(1-2\nu)(d\,x^2 + a\,y^2)\right)$$
(C.3b)

$$W_{1113}(x,y) = \frac{Q}{2\pi\varepsilon_0(\kappa+1)} \frac{1+\nu}{2\pi Y} \frac{\pi x}{4a(a+d)^2 \rho^2} \times$$
$$\times \left((5a+2d)x^2 + (-a+8d)y^2 + 2(1-2\nu)((a+2d)x^2 + 3a\,y^2)\right)$$
(C.3c)

$$W_{1121}(x,y) = \frac{Q}{2\pi\varepsilon_0(\kappa+1)} \frac{1+\nu}{2\pi Y} \frac{\pi x y}{2a(a+d)^3 \rho^2} \times$$
$$\times \left((2a-d)x^2 + (-a+2d)y^2 + 2(1-2\nu)(d\,x^2 + a\,y^2)\right)$$
(C.3d)

$$W_{1122}(x,y) = -\frac{Q}{2\pi\varepsilon_0(\kappa+1)}\frac{1+\nu}{2\pi Y}\frac{\pi}{4a(a+d)^2\rho^4} \times$$
$$\times \begin{pmatrix} a(4a+d)x^4 - 2(a^2 - 7ad + d^2)x^2y^2 + d(a+4d)y^4 + \\ + 2(1-2\nu)(ad\,x^4 + 2(a^2 - ad + d^2)x^2y^2 + ad\,y^4) \end{pmatrix} \quad \text{(C.3e)}$$

$$W_{1123}(x,y) = \frac{Q}{2\pi\varepsilon_0(\kappa+1)}\frac{1+\nu}{2\pi Y}\frac{\pi y}{4a(a+d)^2\rho^2} \times$$
$$\times \left((7a-2d)x^2 + (a+4d)y^2 + 2(1-2\nu)\left((-a+2d)x^2 + a y^2\right)\right) \quad \text{(C.3f)}$$

$$W_{1221}(x,y) = \frac{Q}{2\pi\varepsilon_0(\kappa+1)}\frac{1+\nu}{2\pi Y}\frac{\pi\left(ad\,x^4 + 2(a^2-ad+d^2)x^2y^2 + ad\,y^4\right)}{4a(a+d)^2\rho^4}(1+4\nu) \quad \text{(C.3g)}$$

$$W_{1222}(x,y) = -\frac{Q}{2\pi\varepsilon_0(\kappa+1)}\frac{1+\nu}{2\pi Y}\frac{\pi x y(a x^2 + d y^2)}{2a(a+d)^3\rho^2}(1+4\nu) \quad \text{(C.3h)}$$

$$W_{1223}(x,y) = -\frac{Q}{2\pi\varepsilon_0(\kappa+1)}\frac{1+\nu}{2\pi Y}\frac{\pi x(a x^2 + (-a+2d)y^2)}{4a(a+d)^2\rho^2}(1+4\nu) \quad \text{(C.3i)}$$

$$W_{1131}(x,y) = -\frac{Q}{2\pi\varepsilon_0(\kappa+1)}\frac{1+\nu}{2\pi Y}\frac{\pi x\left((3a+2d)x^2 + (a+4d)y^2\right)}{4a(a+d)^2\rho^2} \quad \text{(C.3j)}$$

$$W_{1132}(x,y) = -\frac{Q}{2\pi\varepsilon_0(\kappa+1)}\frac{1+\nu}{2\pi Y}\frac{\pi y\left((5a+2d)x^2 + (3a+4d)y^2\right)}{4a(a+d)^2\rho^2} \quad \text{(C.3k)}$$

$$W_{1133}(x,y) = -\frac{Q}{2\pi\varepsilon_0(\kappa+1)}\frac{1+\nu}{2\pi Y}\frac{\pi\left((2a+d)x^2 + (a+2d)y^2\right)}{2a(a+d)\rho^2} \quad \text{(C.3l)}$$

$$W_{1231}(x,y) = \frac{Q}{2\pi\varepsilon_0(\kappa+1)}\frac{1+\nu}{2\pi Y}\frac{\pi y\left((-a+2d)x^2 + a y^2\right)}{4a(a+d)^2\rho^2} \quad \text{(C.3m)}$$

$$W_{1232}(x,y) = \frac{Q}{2\pi\varepsilon_0(\kappa+1)}\frac{1+\nu}{2\pi Y}\frac{\pi x\left(a x^2 + (-a+2d)y^2\right)}{4a(a+d)^2\rho^2} \quad \text{(C.3n)}$$

$$W_{1233}(x,y) = -\frac{Q}{2\pi\varepsilon_0(\kappa+1)}\frac{1+\nu}{2\pi Y}\frac{\pi x y}{2a(a+d)^2} \quad \text{(C.3o)}$$

$$W_{1331}(x,y) = \frac{Q}{2\pi\varepsilon_0(\kappa+1)}\frac{1+\nu}{2\pi Y}\frac{\pi(d x^2 + a y^2)}{2a(a+d)\rho^2}(-1+4\nu) \quad \text{(C.3p)}$$

$$W_{1332}(x,y) = -\frac{Q}{2\pi\varepsilon_0(\kappa+1)}\frac{1+\nu}{2\pi Y}\frac{\pi x y}{2a(a+d)^2}(-1+4\nu) \quad \text{(C.3q)}$$

$$W_{1333}(x,y) = -\frac{Q}{2\pi\varepsilon_0(\kappa+1)}\frac{1+\nu}{2\pi Y}\frac{\pi x}{2a(a+d)}(-1+4\nu) \quad \text{(C.3r)}$$

Components related to $u_2$ can be obtained from (C.3) with the help of the simultaneous permutation of indexes "1" ↔ "2" and coordinates x ↔ y. Components determining the displacement component $u_3$ are the following:

$$W_{3111}(x,y) = \frac{Q}{2\pi\varepsilon_0(\kappa+1)} \frac{1+\nu}{2\pi Y} \frac{\pi x\left((a+2d)x^2+3ay^2\right)}{4a(a+d)^2\rho^2}(-1+4\nu) \quad \text{(C.4a)}$$

$$W_{3112}(x,y) = \frac{Q}{2\pi\varepsilon_0(\kappa+1)} \frac{1+\nu}{2\pi Y} \frac{\pi y\left((-a+2d)x^2+ay^2\right)}{4a(a+d)^2\rho^2}(-1+4\nu) \quad \text{(C.4b)}$$

$$W_{3113}(x,y) = \frac{Q}{2\pi\varepsilon_0(\kappa+1)} \frac{1+\nu}{2\pi Y} \frac{\pi\left(dx^2+ay^2\right)}{2a(a+d)\rho^2}(-1+4\nu) \quad \text{(C.4c)}$$

$$W_{3121}(x,y) \equiv W_{3112}(x,y) \quad \text{(C.4b)}$$

$$W_{3123}(x,y) = -\frac{Q}{2\pi\varepsilon_0(\kappa+1)} \frac{1+\nu}{2\pi Y} \frac{\pi xy}{2a(a+d)^2}(-1+4\nu) \quad \text{(C.4d)}$$

$$W_{3131}(x,y) = -\frac{Q}{2\pi\varepsilon_0(\kappa+1)} \frac{1+\nu}{2\pi Y} \frac{\pi\left(dx^2+ay^2\right)}{2a(a+d)\rho^2} \quad \text{(C.4e)}$$

$$W_{3132}(x,y) = \frac{Q}{2\pi\varepsilon_0(\kappa+1)} \frac{1+\nu}{2\pi Y} \frac{\pi xy}{2a(a+d)^2} \quad \text{(C.4f)}$$

$$W_{3133}(x,y) = \frac{Q}{2\pi\varepsilon_0(\kappa+1)} \frac{1+\nu}{2\pi Y} \frac{\pi x}{2a(a+d)} \quad \text{(C.4g)}$$

$$W_{3331}(x,y) = -\frac{Q}{2\pi\varepsilon_0(\kappa+1)} \frac{1+\nu}{2\pi Y} \frac{\pi x}{2a(a+d)}(3-4\nu) \quad \text{(C.4h)}$$

$$W_{3333}(x,y) = -\frac{Q}{2\pi\varepsilon_0(\kappa+1)} \frac{1+\nu}{2\pi Y} \frac{\pi}{2a}(3-4\nu) \quad \text{(C.4i)}$$

Here we listed only those components that cannot be found with the help of the above-mentioned rule.

### APPENDIX D. Integration on $\varphi$.

The integration on polar angle $\varphi$ reduces to the following integrals

$$\int_0^{2\pi} \frac{d\varphi}{\alpha+\beta\cos(\varphi)+\chi\sin(\varphi)} = \frac{2\pi}{\sqrt{\alpha^2-\beta^2-\chi^2}} \quad \text{(D.1)}$$

$$\int_0^{2\pi} \frac{\cos(\varphi)d\varphi}{\alpha+\beta\cos(\varphi)+\chi\sin(\varphi)} = -\frac{2\pi\beta}{\sqrt{\alpha^2-\beta^2-\chi^2}} \frac{1}{\sqrt{\alpha^2-\beta^2-\chi^2}+\alpha} \quad \text{(D.2)}$$

$$\int_0^{2\pi} \frac{\sin(\varphi)d\varphi}{\alpha + \beta\cos(\varphi) + \chi\sin(\varphi)} = -\frac{2\pi\chi}{\sqrt{\alpha^2 - \beta^2 - \chi^2}} \frac{1}{\sqrt{\alpha^2 - \beta^2 - \chi^2} + \alpha} \quad (D.3)$$

$$\int_0^{2\pi} \frac{\cos(\varphi)^2 d\varphi}{\alpha + \beta\cos(\varphi) + \chi\sin(\varphi)} = \frac{2\pi}{(\beta^2 + \chi^2)^2}\left(-\alpha(\beta^2 - \chi^2) + \frac{\alpha^2(\beta^2 - \chi^2) + \chi^2(\beta^2 + \chi^2)}{\sqrt{\alpha^2 - \beta^2 - \chi^2}}\right) \quad (D.4)$$

$$\int_0^{2\pi} \frac{\sin(\varphi)^2 d\varphi}{\alpha + \beta\cos(\varphi) + \chi\sin(\varphi)} = \frac{2\pi}{(\beta^2 + \chi^2)^2}\left(\alpha(\beta^2 - \chi^2) + \frac{-\alpha^2(\beta^2 - \chi^2) + \beta^2(\beta^2 + \chi^2)}{\sqrt{\alpha^2 - \beta^2 - \chi^2}}\right) \quad (D.5)$$

$$\int_0^{2\pi} \frac{\cos(\varphi)\sin(\varphi)d\varphi}{\alpha + \beta\cos(\varphi) + \chi\sin(\varphi)} = \frac{2\pi\beta\chi}{2\alpha(\alpha^2 - \beta^2 - \chi^2) + (2\alpha^2 - \beta^2 - \chi^2)\sqrt{\alpha^2 - \beta^2 - \chi^2}} \quad (D.6)$$

$$\int_0^{2\pi} \frac{\sin(\varphi)^3 d\varphi}{\alpha + \beta\cos(\varphi) + \chi\sin(\varphi)} =$$
$$= \frac{2\pi\chi}{(\beta^2 + \chi^2)^3}\left(-2\alpha^2(3\beta^2 - \chi^2) + (3\beta^2 + \chi^2)(\beta^2 + \chi^2) + \frac{\alpha^3(6\beta^2 - 2\chi^2) - 6\alpha\beta^2(\beta^2 + \chi^2)}{\sqrt{\alpha^2 - \beta^2 - \chi^2}}\right) \quad (D.7)$$

$$\int_0^{2\pi} \frac{\cos(\varphi)^3 d\varphi}{\alpha + \beta\cos(\varphi) + \chi\sin(\varphi)} =$$
$$= \frac{2\pi\beta}{(\beta^2 + \chi^2)^3}\left(2\alpha^2(\beta^2 - 3\chi^2) + (\beta^2 + 3\chi^2)(\beta^2 + \chi^2) + \frac{-\alpha^3(2\beta^2 - 6\chi^2) - 6\alpha\chi^2(\beta^2 + \chi^2)}{\sqrt{\alpha^2 - \beta^2 - \chi^2}}\right) \quad (D.8)$$

It should noted that (D.1)-(D.8) are taken under conditions $\alpha > 0$, $\alpha^2 > \beta^2 + \chi^2$ that is indeed true since $\alpha \equiv d$, $\beta \equiv ix$, $\chi \equiv iy$.